\documentclass[12pt,twoside,letterpaper]{article} %% The same for {book}
\usepackage{times,fancyhdr}
\usepackage[dvips]{graphicx}

\usepackage{amsmath}
\usepackage{amsfonts}
	
\usepackage{epstopdf}	

\usepackage{amssymb}
\usepackage{times,fancyhdr}
\begin{document}

%............................ definitions ..............
\newcommand{\be}{\begin{equation}}
\newcommand{\ee}{\end{equation}}
\def\bq{\begin{eqnarray}}
\def\eq{\end{eqnarray}}
%........................................................

\title{{\bfseries  Resolving Hubble tension with the Milne model\footnote{Received an {\sc Honorable Mention}  in the 2020 Essay Competition of the Gravity Research Foundation.}}}
\author{\bfseries\itshape Ram Gopal Vishwakarma\\
 Unidad Acad$\acute{e}$mica de Matem$\acute{a}$ticas\\
 Universidad Aut$\acute{o}$noma de Zacatecas,\\
  Zacatecas C. P. 98068, ZAC, Mexico\\
vishwa@uaz.edu.mx
}

\date{}
\maketitle
\thispagestyle{empty}
\setcounter{page}{1}

%\title{\bf Resolving Hubble tension with the Milne model\footnote{{Received an {\sc Honorable Mention}  in the 2020 Essay Competition of the Gravity Research Foundation.}}}

\begin{abstract}
The recent measurements of the Hubble constant based on the standard $\Lambda$CDM cosmology reveal an underlying disagreement between the early-Universe estimates and the late-time measurements. Moreover, as these measurements improve, the discrepancy not only persists but becomes even more  significant and harder to ignore. The present situation places the standard cosmology in jeopardy and provides a tantalizing hint that the  problem results from some new physics beyond the  $\Lambda$CDM model.

It is shown that a non-conventional theory - the Milne model - which  introduces a different evolution dynamics for the Universe, alleviates the Hubble tension significantly. Moreover, the model also averts some long-standing problems of the standard cosmology, for instance, the problems related with the cosmological constant, the horizon, the flatness, the Big Bang singularity, the age of the Universe and the non-conservation of energy.
\\


\end{abstract}

% ------------ [Running Heads - for odd and even pages] - please insert it only on page 2!
\pagestyle{fancy}
\fancyhead{}
\fancyhead[EC]{\it R. G. Vishwakarma}
\fancyhead[EL,OR]{\thepage}
\fancyhead[OC]{\it Resolving Hubble tension with the Milne model}
\fancyfoot{}
\renewcommand\headrulewidth{0.5pt}
%------------------------------------------------------------------------------

\section{Introduction}

The Hubble constant $H_0$ is one of the most important parameters in cosmology which measures the present rate of  expansion of the Universe, and thereby estimates its size and age.  This is done by assuming a cosmological model. However, the two values of $H_0$ predicted by the standard $\Lambda$CDM model, one inferred from the measurements of the early Universe and the other from the late-time local measurements, seem to be in sever tension.
As the two values are measured with greater and greater precision, the discrepancy becomes stronger and stronger.
While fitting the $\Lambda$CDM model to the cosmic microwave background (CMB) data obtained by the Planck satellite gives $H_0= 67.4  \pm0.5$ km/s/Mpc \cite{Planck}, the fitting of this model to the Cepheids-calibrated supernovae of Type Ia (SNe Ia) data from the SHOES project gives $H_0 = 74.03 \pm1.42$ km/s/Mpc \cite{Shoes}, which is at 4.4 $\sigma$ discrepancy compared
with that inferred from the CMB.
As this discrepancy is not attributable to any unaccounted systematic effects in the data, it appears to result from a cosmological feature beyond $\Lambda$CDM \cite{Shoes}.

Although a large number of mechanisms have been proposed in order
to resolve this tension, none of these suggestions is compelling. Given the challenge of this
problem, it is worthwhile to explore models beyond the $\Lambda$CDM and standard General Relativity (GR). We consider, for this purpose, a less conventional theory (but still sharing many features of the standard cosmology) - the Milne model - which was proposed by Edward Arthur Milne way back in 1935 \cite{milne}.
Although the model requires physics beyond the $\Lambda$CDM,  it
 alleviates the $H_0$ tension significantly, besides evading  many other problems of the standard cosmology, as we shall see in the following.
As Milne's model is not a widely known theory, we describe it briefly in the following (a good review of the theory can be found in Ref. \cite{Bondi}).

\section{Milne's Theory}

There is a widespread misunderstanding about the Milne model that persists in the literature. With the opt abused argument that the Milne model is a specific GR model - the {\it empty} Universe, the model is considered nonphysical and hence is ignored, which is though misleading. It should be noted that the Milne model results, independently  of GR, as a natural consequence of kinematic relativity and cosmological principle.

 Kinematic relativity is a deductive theory which attempts to deduce a fact from the observables and already known  ordinary physics - here the cosmological principle and the basic properties of spacetime and the propagation of light.
As the curved spacetime is never directly observed, Milne attempted a simpler treatment of the world structure by arguing that a consistent model of the Universe should be subject to Special Relativity (SR), because the spacetime appears flat.
The motivation was to have the simple ordinary physics of flat static spacetime, evading the intervening calculations of curved Riemannian spacetime. 

As has been mentioned above, Milne's theory is based on the cosmological principle - that all the observers, irrespective of their position and state of motion, obtain the same aspect of the Universe.
In order to realize this principle, it becomes necessary to select a set of homogeneously distributed `fundamental particles and observers' of whom there is always only one at each point. 
This is equivalent to Weyl's bundle of geodesics with respect to which the cosmological principle is introduced into standard cosmology. Milne immediately identified the members of this substratum - the system of fundamental particles - with the galaxies and the fundamental observers moving with the galaxies. This plays the role of a homogeneous background, against which the local inhomogeneities and the random motions have to be considered. 

It has already been shown by Robertson \cite{Robertson} and Walker \cite{Walker}, by employing the methods of the theory of groups, that the most general line element satisfying the conditions of isotropy and homogeneity required by the cosmological principle, is
\be
ds^2=c^2 d\bar{t}^2-S^2(\bar{t})\left[\frac{dr^2}{1-k r^2}+r^2(d\theta^2+\sin^2\theta ~d\phi^2)\right],\label{eq:RW}
\ee
which is now known as the Robertson-Walker (RW) line element. By the use of the transformations  $r=\frac{\bar{r}}{1+k\bar{r}^2/4},  \bar{x}=\bar{r}\sin\theta\cos\phi, \bar{y}=\bar{r}\sin\theta\sin\phi$ and $\bar{z}=\bar{r}\cos\theta$, this metric can be recast in its conformally flat forms
\be
ds^2=c^2 d\bar{t}^2-S^2(\bar{t})\left[\frac{d\bar{r}^2+\bar{r}^2(d\theta^2+\sin^2\theta ~d\phi^2)}{(1+\frac{k}{4} \bar{r}^2)^2}\right]=c^2 d\bar{t}^2-S^2(\bar{t})\frac{d\bar{x}^2+d\bar{y}^2+d\bar{z}^2}{[1+\frac{k}{4}( \bar{x}^2 +\bar{y}^2 +\bar{z}^2)]^2}.\label{eq:RWcon}
\ee
Milne considered the scale factor $S=c\bar{t}$ in the line element in order to make the motion of the observers uniform. Now, $k=-1$ remains the only choice to make the RW line element compatible with the Minkowskian metric, since with $S=c\bar{t}$, the resulting 4-dimensional spacetime  is flat only when $k=-1$ and the 3-space is hyperbolic. The identification
\be
S=c\bar{t}, ~~~~~k=-1\label{eq:milneSol}
\ee
reduces the RW metric to the form
\be
ds^2=c^2 d\bar{t}^2-c^2\bar{t}^2\left[\frac{d\bar{r}^2+\bar{r}^2(d\theta^2+\sin^2\theta ~d\phi^2)}{(1- \bar{r}^2/4)^2}\right],\label{eq:milne}
\end{equation}
which is in fact Minkowskian as we shall see in the following.

Let us note that the Einstein equation $R^{\mu\nu}-\frac{1}{2}R^\sigma_\sigma g^{\mu\nu}=-\frac{8\pi G}{c^4}T^{\mu\nu}$, when solved for the RW metric, provides the Friedmann equation 
\be
\frac{\dot{S}^2}{S^2}+\frac{kc^2}{S^2}=\frac{8\pi G\epsilon}{3},\label{eq:friedmann}
\ee
which gives the evolution of the energy density $\epsilon$ in the standard cosmology. As this equation leads to the same Milne solution (\ref{eq:milneSol}) for a vanishing energy density, it is (erroneously) believed that the empty FLRW model is the Milne model, despite the glaring fact that Milne has identified the fundamental particles of the Universe with the very galaxies, as mentioned above.
This readily demonstrates that the Milne model results from a gravitational theory other than the one governed by Einstein equation with a non-vanishing $T^{\mu\nu}$.   Clearly, the Milne's theory is not a particular case of GR and hence it cannot be recast in the framework of GR. 
All one can say, in the language of GR, is that matter does not curve the spacetime in the GR-analogue of Milne model.
The presence of matter without curving spacetime\footnote{The appearance of a flat spacetime in 
the presence of matter is also not impossible in the conventional GR. 
For example, it has been shown in Ref. \cite{Ayon-Beato} that 
conformally coupled matter does not always curve the spacetime.} in Milne's theory, indicates that this theory is fundamentally different from GR and should not be viewed within the usual understanding of an empty Universe of GR. 
It may also be noted that the Milne's theory must also be different from the Newtonian gravitation considered in a cosmological scenario, since equation (\ref{eq:friedmann}) for dust can also be derived in the Newtonian theory \cite{narlikar}, and we have already seen that equation (\ref{eq:friedmann}) is not compatible with Milne's model.

An illuminating achievement of Milne's kinematic relativity is the possibility of the existence of multiple time scales. One such important scale is the cosmic time scale $\bar{t}$ used in the RW metric (\ref{eq:milne}), in which the relative motion of the observes is non-vanishing but unaccelerated (as it is a special-relativistic theory). Another important time scale, say $\tau$,  uses a local time in which the observers appear to be at rest and the Universe presents a static appearance.
The $\tau$-time is related with the $\bar{t}$-time through the transformation
\be
\tau=t_0 \ln\left(\frac{\bar{t}}{t_0} \right),\label{eq:trans}
\ee
which transforms the line element (\ref{eq:milne}) to a form conformal to a static form of (\ref{eq:milne}):
\be
ds^2=e^{2\tau/t_0}\left[c^2 d\tau^2-c^2t_0^2\left\{\frac{d\bar{r}^2+\bar{r}^2(d\theta^2+\sin^2\theta ~d\phi^2)}{(1-\bar{r}^2/4)^2}\right\}\right].\label{eq:milnen}
\ee
Here $t_0$ is a constant with the significance that $\tau=0$ when $\bar{t}=t_0$. While the line element (\ref{eq:milne}) uses the comoving coordinates and a cosmic time, the line element (\ref{eq:milnen}) uses the locally defined measures of space and time.
The zero of $\bar{t}$-time scale is a fundamental event in the theory when the separation of the fundamental (co-moving) observers vanishes, with all the matter in the Universe  accumulated at a single spatial point proposing a physical explosion of matter. In $\tau$-time scale, this event takes place in the infinite past, by virtue of its logarithmic dependence on $\bar{t}$, as is indicated by (\ref{eq:trans}).

In order to complete the discussion of the substratum in Milne's model, we need to determine the evolution of the energy density. For this purpose, we consider Milne's preferred $t$-time scale together with the 3-space Euclidean. The required transformations are
\be
t=\bar{t}\left[\frac{1+\bar{r}^2/4}{1-\bar{r}^2/4}\right],x=\frac{c\bar{x}\bar{t}}{\left(1-\bar{r}^2/4\right)},y=\frac{c\bar{y}\bar{t}}{\left(1-\bar{r}^2/4\right)},z=\frac{c\bar{z}\bar{t}}{\left(1-\bar{r}^2/4\right)}~~\text{with}~~ \bar{r}^2=\bar{x}^2+\bar{y}^2+\bar{z}^2,\label{eq:transn}
\ee
which transform the line element (\ref{eq:milne}) to the canonical form
\be
ds^2=c^2dt^2- dx^2- dy^2- dz^2,\label{eq:milnenn}
\ee 
which uses the locally defined measures of time and space in $t$-time. By recalling that the comoving coordinate $\bar{r}$ does not change for any observer along its geodesic, one can calculate the transformation from $t$-time to $\tau$-time from equations (\ref{eq:trans}) and (\ref{eq:transn}), giving
\be
\tau=t_0 \ln\left( t/t_0 \right) + t_0 ,\label{eq:ttime}
\ee
with a suitable shift in the origin of the $\tau$-time scale. In this way, $t$-time and $\tau$-time agree both in value and rate at $t=t_0$. Like $\bar{t}=0$, the event $t=0$ too takes place in the infinite past in $\tau$-time scale. Milne preferred to use $t$-time as his fundamental scale. Now we can derive the variation of density with $t$-time as the following.

As we have noted earlier, the coordinates $\bar{r},\bar{x},\bar{y},\bar{z}$ are constant for any observer. It then follows from the transformations (\ref{eq:transn}) that $x/t$, $y/t$ and $z/t$ are constant, determining a simple velocity-distance relation as
\be
v_x = \frac{dx}{dt}=\frac{x}{t}, ~~~~v_y = \frac{dy}{dt}=\frac{y}{t}, ~~~~v_z = \frac{dz}{dt}=\frac{z}{t}.\label{eq:vel}
\ee
We also note that besides $ds$, there is another quantity
\be
c^2t^2-x^2-y^2-z^2 \equiv X,
\ee
which is invariant (independent of the observer), as is implied by the flat spacetime and the Lorentz transformations. 
By virtue of this and relations  (\ref{eq:vel}), equation (\ref{eq:milnenn}) then provides
\be
\left(\frac{ds}{dt}\right)^2=c^2 - \left(\frac{dx}{dt}\right)^2 - \left(\frac{dy}{dt}\right)^2 - \left(\frac{dz}{dt}\right)^2 = \frac{X}{t^2}.\label{eq:ds/dt}
\ee
One expects that the density $\rho$ measured in the $(t,x,y,z)$ coordinate system must also be an invariant since it must present the same aspect to every fundamental observer, as is expected from the cosmological principle.
Thus $\rho=$ the number of particles in the invariant volume element $dx\,dy\,dz\,dt/ds$, can only be a function of $X$. Hence the apparent density $n=$
the number of particles in the volume element $dx\,dy\,dz$ is given by
\be
n = \rho(X) \frac{dt}{ds} = \frac{t \rho(X)}{\sqrt{X}},\label{eq:n}
\ee
by virtue of (\ref{eq:ds/dt}). 
Let us now make a plausible assumption that matter is conserved and hence the equation of hydrodynamic continuity
\be
\frac{\partial n}{\partial t} + \vec{\nabla}\cdot(n \vec{v})=0, \label{eq:conti}
\ee
applies. Since no random motions have been introduced, the question of pressure does not arise. The use of equations (\ref{eq:vel}) and (\ref{eq:n}) in (\ref{eq:conti}), and some simple calculations, lead to
\be
\rho = A X^{-3/2}, ~~~ n = At X^{-2}, ~~~ A = \text{constant}.\label{eq:rho}
\ee
Thus  the density of matter decreases with time in the Universe whose invariant border $X=0$ advances at the speed of light.
Further, the density increases from the origin towards $X=0$.

\section{SNe Ia in the Robertson-Walker Models}

We have mentioned that a lower value for the Hubble parameter $H_0$ is measured from the CMB observations.
Let us however note that this low value of $H_0$ is also corroborated by other independent measurements, for instance the BAO data combined with either CMB power spectra or deuterium abundance measurements \cite{Addition}. Thus, it is the (higher) value of $H_0$ inferred from the Cepheids-calibrated SNe Ia in the framework of $\Lambda$CDM model, which remains to be examined.
Additionally, the observations on the SNe Ia  are the most important and decisive ones, among the various late-time cosmological observations, and hence need attention in view of the Hubble tension.  

It  has  already  been noticed that  the  coasting  model (\ref{eq:milne}), i.e. the RW metric (\ref{eq:RW})  taken together with the Milne solution (\ref{eq:milneSol}),  is  consistent  with  the  observations  of SNe Ia without requiring any dark energy.
As early as in 1999, the Supernova Cosmology Project team realized from the analysis of their first-generation of the SNe Ia data that the performance of the 
model (\ref{eq:milne})  is practically identical to that of the best-fit $\Lambda$CDM model \cite{Perlmutter}. Since then, the model (\ref{eq:milne}) has been shown to fare well with not only the new SNe Ia data, but also other observations \cite{PhysScripta, Universe}.

In order to fit a model to the SNe Ia data, one notes that the models based on the line-element (\ref{eq:RW}), predict that the observed magnitude $m$ of an SN Ia at a distance $d _
{\mbox{{\scriptsize L}}}$ at redshift $z$ is given by
\begin{equation}
 m(z,\Omega_i)={\cal M} +
5 \log\left(\frac{H_0}{c} d _
{\mbox{{\scriptsize L}}}(z,\Omega_i)
\right),\label{eq:m-z}
\end{equation}
where $\Omega_i$ are the adjustable parameters, if any, of the considered model and 
\be
{\cal M} \equiv M - 5 \,\log H_0 +  {\rm constant},  \label{eq:M-H}
\ee
with $M$ being the absolute magnitude of the considered SN Ia.
The value of the `constant' appearing in  (\ref{eq:M-H}) depends on the chosen units in which
$d_{\rm L}$ and $H_0$ are measured. It may be mentioned that
the zero-point absolute magnitudes of SNe Ia are generally set arbitrarily in different data sets. While fitting the combined
data, this situation is handled successfully by this `constant'  which now
plays the role of the normalization constant and simply gets modified suitably. It may also be mentioned that sometimes the data are given in terms of the distance modulus 
$\mu\equiv m-M$, instead of the magnitude $m$. In this case, $m$ is replaced by $\mu$ in equation (\ref{eq:m-z}) and ${\cal M}$ is given by ${\cal M}={\rm constant}- 5 \,\log H_0$.  

Interestingly the distance $d _{\mbox{{\scriptsize L}}}$, in the Milne model, is completely determined by the solution (\ref{eq:milne}), without requiring any input from
the matter fields: 
$
d_{\rm L}=cH_0^{-1}z(2+z)/2.
$
 On the other hand, the contributions from the matter fields, entering through the Friedmann equation (\ref{eq:friedmann}), determine $d _{\mbox{{\scriptsize L}}}$  in the standard cosmology.

Equation (\ref{eq:M-H}) indicates that a precise measurements of $H_0$ requires the calibration of the absolute supernova magnitude $M$. However, this can be avoided and one can still estimate $H_0$ in a model, in terms of $H_0$ estimated in another model for instance the $\Lambda$CDM  model, as is directed by equation (\ref{eq:M-H}) giving
\be
\frac{H_0^{\mbox{{\scriptsize M}}}}{H_0^{\mbox{{\scriptsize $\Lambda$}}}} = 10^{({\cal M}^{\mbox{{\scriptsize $\Lambda$}}} - {\cal M}^{\mbox{{\scriptsize M}}})/5},  \label{eq:H0}
\ee
which is even more informative and directly compares the relative performance of the models regarding the estimation of $H_0$ from an SNe Ia data set.
Here the superscripts  ${\mbox{{\scriptsize M}}}$ and ${\mbox{{\scriptsize $\Lambda$}}}$ on a parameter denote the value of the parameter estimated from the considered SNe Ia data in  respectively the Milne model and the $\Lambda$CDM model.

Let us now consider the Pantheon compilation of SNe Ia which consists of 1048 points in the range $0.01 < z< 2.26$  calibrated with the SHOES distance ladder
 \cite{Pantheon}. The compilation provides the biggest and the newest SNe Ia sample publicly available today.
Fitting the $\Lambda$CDM model to this sample gives $\Omega_{\mbox{{\scriptsize m}}}=1-\Omega_{\mbox{{\scriptsize $\Lambda$}}}=0.2852 \pm 0.0002$ with  ${\cal M}^{\mbox{{\scriptsize $\Lambda$}}}=23.8036$ at $\chi^2/{\rm DoF}=0.99$ as the best-fitting solution (DoF $\equiv$ Degrees of Freedom). Similarly, fitting the Milne model to the sample gives the value of the only free parameter ${\cal M}^{\mbox{{\scriptsize M}}}=23.8886$ at $\chi^2/{\rm DoF}=1.10$ as the best-fitting solution.

The use of these values of ${\cal M}$ in equation  (\ref{eq:H0}) estimates the ratio $H_0^{\mbox{{\scriptsize M}}}/H_0^{\mbox{{\scriptsize $\Lambda$}}}=0.9616$, a value lower than 1 and hence giving a smaller $H_0$ in the Milne model compared to the corresponding $H_0$ in the $\Lambda$CDM model. For instance, the values of $H_0$ in the $\Lambda$CDM model estimated from the Pantheon sample in Ref. \cite{Arman} are 72.61 km/s/Mpc and 71.19 km/s/Mpc at $1\sigma$ and $2\sigma$ respectively. This leads to the corresponding values of $H_0$ in the Milne model as 69.82 km/s/Mpc  and 68.46 km/s/Mpc respectively.
In view of $H_0= 67.4$ km/s/Mpc estimated from the CMB, these values estimated in the Milne model alleviate the $H_0$-tension significantly, if not removing it completely.

\begin{table}
\caption{\bf $H_0^{\mbox{{\scriptsize M}}}/H_0^{\mbox{{\scriptsize $\Lambda$}}}$ estimated from different SNe Ia Samples (the JLA sample has not been included in the analysis as it shows some inconsistencies in the correction for peculiar velocities \cite{Sarkar}). The full set of parameters of the $\Lambda$CDM model estimated from the data is not shown here.} 
\begin{center}
    \begin{tabular}{ | l || l | l | l |}
    \hline
 ~~~~~~~~Sample Name \& Year   &~~~~Flat $\Lambda$CDM & ~~~~~~~~Milne&       
$H_0^{\mbox{{\scriptsize M}}}/H_0^{\mbox{{\scriptsize $\Lambda$}}}$\\

&$\chi^2/{\rm DoF}$  ~~~ ${\cal M}^{\mbox{{\scriptsize $\Lambda$}}}$ &$\chi^2/{\rm DoF}$ ~~~ ${\cal M}^{\mbox{{\scriptsize M}}}$&\\
\hline 
`Gold'  (157 SNe Ia) 2004 \cite{gold} & ~1.14 ~~~~~ 43.3423 & ~1.23 ~~~~~ 43.3984 & ~~0.9745 \\  [2ex]

`New Gold'  (182 SNe Ia) 2007  \cite{Riess}& ~0.88 ~~~~~ 43.3957 & ~0.96 ~~~~~ 43.4525 & ~~0.9742 \\  [2ex]

`Union 2.1' (580 SNe Ia) 2012\cite{Union} & ~0.97 ~~~~~ 43.1585 & ~1.04 ~~~~~ 43.2365 & ~~0.9647 \\  [2ex]

`Pantheon' (1048 SNe Ia) 2018\cite{Pantheon}    & ~0.99 ~~~~~ 23.8036 & ~1.10 ~~~~~ 23.8886 & ~~0.9616\\  [2ex]

\hline 
    \end{tabular}
\end{center}
\end{table}

We perform this exercise for different SNe Ia samples and estimate the ratio $H_0^{\mbox{{\scriptsize M}}}/H_0^{\mbox{{\scriptsize $\Lambda$}}}$. The results are shown in Table 1. It is clear from the table that as  more and more SNe Ia with higher precision are included in the sample, the ratio $H_0^{\mbox{{\scriptsize M}}}/H_0^{\mbox{{\scriptsize $\Lambda$}}}$ becomes smaller and smaller. From this trend of the data, we expect an even lower value of the ratio (compared with that obtained for the Pantheon sample) for the SNe Ia data used in the SHOES project \cite{Shoes}. Once this data is made public, a similar exercise can be performed.  However we check, for curiosity, that even with the value of the ratio $H_0^{\mbox{{\scriptsize M}}}/H_0^{\mbox{{\scriptsize $\Lambda$}}}=0.9616$ calculated for the Pantheon data, the measurement  $H_0^{\mbox{{\scriptsize $\Lambda$}}} = 74.03$  by the SHOES project provides $H_0^{\mbox{{\scriptsize M}}}=71.19$, which still alleviates the tension considerably.

It may be mentioned that although the Milne model has a reasonably good fit to the different SNe Ia samples (as is clear from Tablel 1), however the $\Lambda$CDM model shows even a better fit. This has two reasons: First, the $\Lambda$CDM model has more free adjustable parameters. Second, the flat $\Lambda$CDM model itself is assumed in order to calibrate the parameters of the SNe Ia in the standardization process.
Obviously the so calibrated data has to favour the $\Lambda$CDM model then \cite{Universe}.

\section{Other remarkable features of Milne model}

Let us now witness some successes of the Milne model registered on the theoretical front. As we shall see,
the model circumvents the long-standing problems of the standard cosmology, for example, the horizon, the flatness and the cosmological constant problems. It also averts the singularity problem, the problem of the age of the Universe and that of the non-conservation of energy.

\subsection{No Dark Energy, Horizon and Flatness Problems}

The standard cosmology assumes the existence of dark matter, dark energy and inflation in order to explain different observations. However, there is, until now, no non-gravitational or laboratory
evidence for any of these dark sectors.  Additionally, the most favored candidate of dark energy - the cosmological
constant $\Lambda$ - poses a serious confrontation between fundamental physics and standard cosmology.

As all the candidates of dark energy can be assimilated in the energy-stress tensor, and since the latter is absent from the dynamical equations in the Milne theory, the dark energy and its associated problems, for instance the  cosmological constant problem and the coincidence problem, are evaded in the Milne model. We have already seen that the SNe Ia observations are successfully explained in the Milne model without requiring an accelerated expansion or dark energy. Other observations are also explained successfully in the Milne model  \cite{PhysScripta}.

The observed isotropy of CMB cannot be explained in the standard cosmology in terms of some homogenization process taken place in the baryon-photon plasma operating under the principle of causality, since a finite value for the particle horizon    
\[
d_{\rm PH}(\bar{t})=c S(\bar{t}) \int^{\bar{t}}_0 dt'/S(t')
\]
(which is calculated from the RW metric) exists in the theory. As $d_{\rm PH}=\infty$ always for  $S=c\bar{t}$, no horizon exists  in the Milne model and the whole Universe is always causally connected, which explains the observed overall uniformity of CMB without invoking inflation.

The standard cosmology harbours the flatness problem through the Friedmann equation (\ref{eq:friedmann}) which can alternatively be written as
\be
\frac{\epsilon}{3H^2/(8\pi G)} -1  =\Omega-1=\frac{kc^2}{S^2H^2},
\ee
implying that $|\Omega-1|$ grows with time throughout the evolution\footnote{For instance, in the standard cosmology with $S\propto \bar{t}^{1/2}$ in the early Universe, one has $|\Omega-1|\propto \bar{t}$.}. This causes a problem that even a minute departure of early $\Omega$ from unity grows significantly in time  and yet the Universe today remains very close to flat.
For example, the observational uncertainty of $\Omega$ at present, would require it to be differing from unity by $10^{-53}$ during the GUT epoch! Any relaxation of this fine tuning would have led to a far wider range of $\Omega$ at present than is permitted by the observations. 

Let us however note that the evolution of the matter density $\rho$ given by equation (\ref{eq:rho}) in the Milne model does not harbour this problem.

\subsection{No Singularity}

As the Big Bang singularity is a breakdown of the laws of physics and the geometrical structure
of spacetime, there have been attempts to discover singularity-free cosmological solutions of Einstein
equations. Although the line element (\ref{eq:milne}), which represents the cosmological model in Milne's theory, has
well-behaved metric potentials at $\bar{t}= 0$, the volume of the spatial slices vanishes there resulting in
a blowup in the accompanied matter density. However, this is just a coordinate effect since this situation can be averted by transforming the  line element (\ref{eq:milne}) in the manifestly Minkowskian form (\ref{eq:milnenn}).

This also exemplifies that staticity/dynamicity is not an intrinsic property  of a spacetime, but a coordinate-dependent condition. There exist many other cases too when the RW line element reduces to a static form in particularly chosen coordinates \cite{StaticFormRW}. Similar examples exist in other spacetimes also. For instance, the Schwarzschild  spacetime appears static in the usual isotropic Schwarzschild coordinates. But it can be transformed to a dynamic form in another coordinate system \cite{LL}.

\subsection{No Problems with the Age of the Universe and the Conservation of Energy}

It may be noted that the age of the Universe in the $\Lambda$CDM model, when calculated for the value $H_0 = 74.03 \pm1.42$ km s$^{-1}$  Mpc$^{-1}$ from the SHOES project,  comes out as  $12.91 \pm 1.5$ Gyr, which is dangerously close to the age of the oldest globular cluster estimated to be  $t_{\rm GC}=12.5 \pm 1.2$ Gyr \cite{Gnedin} and the age of the Milky Way as $12.5 \pm 3$ Gyr coming from the latest uranium decay estimates \cite{Cayrel}.
Moreover, considering the lower limit of this age of the Universe and the upper limits of the age of the globular cluster and that of the Miky way, the standard cosmology appears in serious trouble.

As has been mentioned earlier, the age of the Universe in the Milne model is infinite in the $\tau$-time scale. 
Interestingly, even in terms of the cosmic time $\bar{t}$, wherein the Universe appears
dynamic in terms of  (\ref{eq:milne}) and the age of the Universe is simply $H_0^{-1}$, this age comes out as 13.73 Gyr which appears satisfactory.

Since the Universe in the Milne model is flat, the symmetries of its Minkowskian form make it
possible to validate the conservation of energy, solving the long-standing problems associated with the
conservation of energy.

\section{Conclusion and Outlook}

To sum up, the Milne's theory provides a simple and lucid structure of the substratum based primarily on the cosmological principle alone. The theory is accredited by various remarkable successes on the theoretical and observational fronts, and supplies a simpler and more elegant model of the Universe than the conventional one.
However, an alternative theory cannot be acceptable purely on the basis of its success on the largest scales. It is also expected to pass the
tests through local observations, for instance those that have been devised to test GR. 
These and many other issues have not been addressed properly in Milne's theory.
Most importantly it lacks a proper theory of gravitation. Let us note that Milne developed his theory without recourse to a specific theory of gravity, be it Einsteinian or Newtonian.
Although he has sketched a law of gravitation, but it is not a general law of gravitation and applies only to mass distributions satisfying the cosmological principle. 
Moreover, none of the several proofs given by Milne in this connection can be considered wholly convincing. 

The historical development of Milne's theory  has left it in a curiously unfinished state, presumably owing to the hostile climate of opinion the theory underwent. It seems that Milne was aware of the shortcomings of his theory, as he has mentioned that his theory is {\it ``similar to that which had been arisen if the general solutions of Kepler's problem had been found from general considerations before the Newtonian dynamics and gravitation had been formulated; ..... or Schwarzschild's form of the metric round a single point-mass had been found before Einstein's field equations were known."}
He hoped that a proper theory may be found later which would embrace his model as a natural consequence. 

It may be mentioned that one such theory can be that which has been formulated recently as a unified theory of gravitation and electrodynamics \cite{IJGMMP}. The theory results from the principles of equivalence and Mach supplemented with a novel insight that the field tensors in a geometric theory of gravitation and electromagnetism must be trace-less, since these long-range interactions are mediated by virtual exchange of massless particles whose mass is expected to be related to the trace of the field tensors. Hence the Riemann tensor, like the analogous electromagnetic field tensor, must be trace-free.
The vanishing of the Ricci tensor then appears as an initial/boundary condition in the theory. Hence, the successes of the classical GR in terms of the Schwarzschild and Kerr etc. solutions, are readily embraced in this theory. 
The structure of spacetime is determined by the energy, momentum and angular momentum (or their densities) of the material plus gravitational fields, which emerge from the geometry itself.
Remarkably, the Milne model naturally emerges as a cosmological solution in the theory when the spacetime becomes homogeneous and isotropic and the positive energy of matter is canceled by the negative gravitational energy.
 Thus the net (material plus gravitational) energy, momentum, angular momentum of the Universe vanishes leading to a vanishing curvature, even though the Universe is not empty.

\vspace{1cm}
\noindent
{\bf Acknowledgements:}
The author thanks an anonymous referee for making some constructive comments, which improved the paper.
The author would also like to thank IUCAA for hospitality, where this work was initiated during a visit.

\end{document}